\documentclass[aps,prb,twocolumn,showpacs,amsmath]{revtex4}

\usepackage{latexsym}
\usepackage{amssymb}
\usepackage{bm}
\usepackage{graphicx}
\usepackage[colorlinks=true, pdfstartview=FitV, linkcolor=blue, 
            citecolor=blue, urlcolor=blue]{hyperref}

\newcommand{\pdag}{{\phantom{\dagger}}}
\newcommand{\bq}{\begin{equation}}
\newcommand{\eq}{\end{equation}}
\newcommand{\bn}{\begin{eqnarray}}
\newcommand{\en}{\end{eqnarray}}

\begin{document}

\title{Shot noise of inelastic tunneling through quantum dot systems}

\author{Bing Dong$^{1,2}$, H. L. Cui$^{1,3}$, X. L. Lei$^{2}$, and Norman J. M. 
Horing$^{1}$} 
\affiliation{$^{1}$Department of Physics and Engineering Physics, Stevens Institute of 
Technology, Hoboken, New Jersey 07030 \\
$^{2}$Department of Physics, Shanghai Jiaotong University,
1954 Huashan Road, Shanghai 200030, China \\
$^{3}$School of Optoelectronics Information Science and Technology, Yantai University, 
Yantai, Shandong, China}

\begin{abstract}

We present a theoretical analysis of the effect of inelastic electron scattering on 
current and its fluctuations in a mesoscopic quantum dot (QD) connected to two leads, 
based on a recently developed nonperturbative technique involving the approximate 
mapping of the many-body electron-phonon coupling problem onto a multichannel 
single-electron scattering problem. In this, we apply the B\"uttiker scattering theory 
of shot noise for a two-terminal mesoscopic device to the multichannel case with 
differing weight factors and examine zero-frequency shot noise for two special cases: 
(i) a single-molecule QD and (ii) coupled semiconductor QDs. The nonequilibrium Green's 
function method facilitates calculation of single-electron transmission and reflection 
amplitudes for inelastic processes under nonequilibrium conditions in the mapping model. 
For the single-molecule QD we find that, in the presence of the electron-phonon 
interaction, both differential conductance and differential shot noise display 
additional peaks as bias-voltage increases due to phonon-assisted processes. In the case 
of coupled QDs, our nonperturbative calculations account for the electron-phonon 
interaction on an equal footing with couplings to the leads, as well as the coupling 
between the two dots. Our results exhibit oscillations in both the current and shot 
noise as functions of the energy difference between the two QDs, resulting from the 
spontaneous emission of phonons in the nonlinear transport process. In the 
``zero-phonon" resonant tunneling regime, the shot noise exhibits a double peak, while 
in the ``one-phonon" region, only a single peak appears.     

\end{abstract}

\maketitle

\section{Introduction}

Recent progress in nanotechnology has made possible the fabrication of single-electron 
tunneling devices using organic molecules, the size of which is sufficiently small so 
that the discrete nature of its energy levels is important. In contrast to semiconductor 
quantum dots (QD), the molecular materials possess much weaker elastic parameters, such 
that it is very easy to excite their internal vibrational degrees of freedom (phonon 
modes) when electrons are incident upon the molecule through a tunnel junction from an 
external environment. This can significantly influence electron tunneling at low 
temperature, and the phenomenon has now provoked a large amount of 
experimental\cite{Chen,Liang,jPark,hPark,Zhitenev} and 
theoretical\cite{Bose,Lundin,Zhu,Flensberg,Braig,Mitra} work on the problem of tunneling 
through a single level with strong coupling to local phonon modes.

On the other hand, phonon-asisted inelastic tunneling in semiconductor QDs has continued 
to attract extensive studies. In particular, a recent experiment has measured the 
nonlinear tunneling current through a tunable two-level quantum system, i.e., coupled 
semiconductor QDs (CQDs), at low temperature with the observation of robust spontaneous 
emission of phonons leading to an oscillatory structure in the current 
spectrum.\cite{Fujisawa} Following this, nonperturbative theoretical analyses ascribed 
this oscillation to an interference effect of the electron-phonon interaction (EPI) in 
CQDs.\cite{Brandes,Keil}   

Thus far, most theoretical work concerning this inelastic tunneling problem has focused 
on calculating current through a two-terminal setup in dc and ac conditions. Only a few 
attempts have been made to study the current-current correlations (shot noise) in QD 
systems,\cite{Zhu,Mitra,Aguado} which has become a central subfield of mesoscopic 
physics.\cite{Buttiker,Blanter,Dong} In the present paper, we investigate low-frequency 
shot noise behavior when the QD is subject to strong coupling with a local phonon mode. 
This kind of strongly nonequilibrium problem has been found to be difficult to solve 
because previous studies demonstrated that when inelastic processes such as phonon 
emission and absorption are considered in the tunneling event through a QD, electron 
scattering becomes a strongly correlated many-body problem involving electrons and the 
various excited phonon states of the 
QD.\cite{Bose,Lundin,Zhu,Flensberg,Braig,Mitra,Brandes,Keil} In this case, the usual 
perturbation theory is invalid for dealing with this scattering problem, albeit that it 
has been extensively exploited with enormous success to study conventional transport in 
bulk materials. 

Recently, a nonperturbative scheme was proposed by Bon\v ca and Trugman\cite{Bonca} to 
study the small polaron of the coupled electron-phonon system described by the Holstein 
model.\cite{Holstein} Later, this method was applied to the analysis of inelastic 
electron scattering in mesoscopic quantum transport in semiconductor QDs and molecular 
wires within the framework of the Landauer-B\"uttiker scattering 
theory.\cite{Bonca2,Ness,Emberly} Very recently, we advantageously employed this mapping 
scheme to evaluate time-dependent phonon-assisted tunneling at low temperature through a 
single-molecular QD with the help of the psuedo-operator technique and the 
nonequilibrium Green's function (NGF) method.\cite{Dongac} The fundamental idea is to 
remodel the strongly correlated Hamiltonian in terms of the combined electron-phonon 
Fock space, mapping the many-body problem onto a one-body scattering problem 
(noninteracting system), which is highly desirable to facilitate an accurate calculation 
of the current-fluctuation spectrum. 

However, this mapping technique introduces an additional difficulty: it also maps the 
ordinary single-mode leads in the two terminals onto pseudo-multi-channel leads 
associated with differing weight factors $P_n$ corresponding to the statistical 
probability of the excited phonon number state $n$ of the mesoscopic device. 
Consequently, the original Landauer current formula needs to be modified to properly 
account for this fact and the Pauli exclusion principle as it applies to multi-channel 
leads.\cite{Bonca2,Emberly,Dongac} For example, if we consider a tunneling event in 
which an electron with energy $\epsilon$ enters from the $n$th channel of the left lead, 
scatters with a phonon in the QD, and finally exits into the $m$th channel of the right 
lead with energy $\epsilon'$, the contribution of this process to current is simply 
proportional to $P_n f_{L}^n(\epsilon) [1-f_{R}^m(\epsilon')]$ ($f_{\eta}^l$ is the 
Fermi distribution function of the $l$th pseudochannel in lead $\eta$, as described in 
Section II below), multiplying the square modulus of the transmission amplitude, 
$|t_{Ln,Rm}|^2$. Correspondingly, it is clear that an analogous modification must be 
made in the B\"uttiker scattering formula for current-current 
correlations\cite{Blanter,Buttiker} in the presence of multichannel leads with differing 
weight factors. This issue is a central focus of this paper. In Section II, we will 
briefly present the appropriate generalization of scattering theory and provide the 
general formula for determining the zero-frequency shot noise spectrum of mesoscopic 
systems with EPI.

In two applications of this generalization we will examine low frequency electric 
current fluctuations as functions of voltage for the two cases mentioned above: 1) a 
single-molecular QD (single site) and 2) semiconductor CQDs (two sites) in the central 
region sandwiched between two normal leads. Every QD is taken to have only one 
single-particle energy level coupled with a dispersionless phonon mode of the 
Holstein-type model. Here we concentrate our attention on the effect of the inelastic 
scattering process in tunneling and ignore all other complexity of real molecular or 
semiconductor devices, except for EPI. For the single site case, our calculations 
predict  (i) a step-type characteristic of the bias voltage-dependent shot noise 
analogous to the current-voltage curve at low temperature, in which the first step 
corresponds to resonance with the single level of the QD and the others result from 
phonon-emission-assisted resonances, and (ii) an enhancement of the Fano factor due to 
phonon effects in the high bias-voltage region. 

For the two-site case, in order to observe the effect of spontaneous phonon emission, we 
consider an extraordinarily high bias-voltage between the two leads and calculate the 
current and shot noise as functions of the energy difference between the two QDs, which 
can be tuned experimentally using gate voltages. Very recently, the shot noise spectrum 
in CQDs was explored by means of master equations with correlation functions in Laplace 
transform space.\cite{Aguado} Under the physical presumption of weak tunneling between 
the QDs and leads, this method employed the noninteracting blip approximation to deal 
with the EPI.\cite{Brandes} Our objection here is to improve upon this by developing an 
approach which allows us to account for the EPI on an equal footing with the couplings 
to the leads, as well as treating the coupling between the two dots in a nonperturbative 
way. The mapping technique conforms to this purpose. Our calculations exhibit an 
oscillatory shoulder in the current and shot noise spectra due to spontaneous phonon 
emission in nonlinear transport. In the resonant tunneling regime, the shot noise has a 
double peak, while only a single peak shows up in the phonon-assisted resonant tunneling 
region.

The rest of this paper is arranged as follows. In Sec. II, we describe the class of 
systems that we study and present the appropriate generalization of scattering theory 
for calculating current fluctuations of a two-terminal device in the presence of EPI. In 
general, we find that the current fluctuation spectrum cannot be expressed in terms of 
transmission and reflection probabilities even in the zero frequency limit, but it is 
sensitive to the transmission and reflection amplitudes involving all pseudochannels.  
Our two applications occupy the following two sections: Sec. III for a single molecular 
QD and Sec. IV for the CQDs. In these two special cases, we employ the NGF technique to 
evaluate transmission and reflection amplitudes in the wide band limit. Numerical 
results and some discussions are also presented in these sections. Finally, our 
conclusions are summarized in Sec. V.     

\section{Method and formulation}

\subsection{Model and mapping Hamiltonian}

Here, we consider the model of the two-terminal mesoscopic device as constituted of a 
central region (the device) connected to two noninteracting reservoirs via tunneling and 
coupled to a phonon bath. We assume that the two leads remain in local equilibrium with 
the Fermi distribution, $f_{\eta}(\omega)=[ 1+ e^{(\omega-\mu_{\eta})/k_{\rm B}T}]^{-1}$ 
[$\mu_{\eta}$ is the chemical potential of the lead $\eta$ ($=L$ or $R$), $T$ is the 
temperature], and both of them have only one electronic channel. We also assume that the 
EPI takes place solely in the device and not in the ideal leads, specifically, there is 
no exchange between electrons and phonons in the tunneling processes. This is reasonable 
because high-order tunneling processes accompanied by phonon emission and absorption are 
much weaker than direct tunneling events. For simplicity, we adopt a single-mode 
Einstein-phonon with frequency $\omega_{ph}$. Moreover, we neglect the spin degree of 
freedom and any effects of electron-electron Coulomb interactions. 

Therefore, the Hamiltonian of this system involves three terms: $H=H_{\rm lead}+H_{\rm 
cen}+H_{\rm tl}$, where $H_{\rm lead}$ describes the two isolated leads, $H_{\rm cen}$ 
models the interacting central region in addition to the free-phonon contribution, and 
$H_{\rm tl}$ is the tunnel coupling between the leads and the device, respectively:  
\begin{subequations}
\label{Hamiltonian}
\bn
H_{\rm lead}&=& \sum_{\eta, k} \epsilon_{\eta k} c_{\eta k}^{\dagger} c_{\eta 
k}^{\pdag}, \label{lead} \\
H_{\rm cen}&=& H_{d}(\{d_{j}^{\dagger}\}; \{d_{j}^{\pdag}\}) + \hbar \omega_{ph} 
b^{\dagger} b^{\pdag} \cr
&& - \sum_{j} \lambda_{j} d_{j}^{\dagger} d_{j}^{\pdag} (b+b^{\dagger}), \label{cen} \\
H_{\rm tl}&=& \sum_{\eta,k,j}V_{\eta,j}(c_{\eta k}^{\dagger} d_{j} + {\rm H.c.}). 
\label{tunneling}
\en    
\end{subequations}
Here $c_{\eta k}^{\dagger}$ ($c_{\eta k}$) are the creation (annihilation) operators for 
the noninteracting electrons with momentum $k$ and energy $\epsilon_{\eta k}$ in the 
lead $\eta$, and $\{d_{j}^{\dagger}\}$ ($\{d_{j}\}$) forms the complete and orthonormal 
set of single-electron creation (annihilation) operators in the central region, 
respectively. The form of $H_{d}$ depends on geometry of the device being investigated. 
The last term in Eq.~(\ref{cen}) denotes the interaction of an electron on the central 
site with phonons: $b^{\dagger}$ ($b$) creates (destroys) a phonon, and $\lambda_{j}$ is 
the on-site EPI constant. $V_{\eta,j}$ stands for the tunnel coupling between the device 
and the lead $\eta$.

Obviously, the model described by the above Hamiltonian (\ref{Hamiltonian}) involves a 
many-body problem with phonon emission and absorption when the electron tunnels through 
the central region. Following the suggestion of Bon\v ca and Trugman, we expand the 
electron states in the central region and the leads in terms of the direct product 
states composed of single-electron states and $n$-phonon Fock states:
\bq
|j,n\rangle=d_j^{\dagger} \frac{(b^{\dagger})^n}{\sqrt{n!}}|0\rangle, \quad (n\geq 0)
\eq
and
\bq
|\eta k,n\rangle=c_{\eta k}^{\dagger} \frac{(b^{\dagger})^n}{\sqrt{n!}}|0\rangle,
\eq
such that the electron state $|j \rangle$ in the central region and the state $|\eta 
k\rangle$ with momentum $k$ in lead $\eta$, respectively, are accompanied by $n$ phonons 
($|0\rangle$ is the vacuum state). After performing this transformation, the many-body 
on-site EPI in Eq.~(\ref{cen}) can be mapped onto a one-body 
model:\cite{Bonca,Bonca2,Ness,Emberly,Dongac}
\bn
\sum_{j,n\geq 0} - \lambda_{j} \sqrt{n+1}(|j,n+1\rangle \langle j,n| + |j,n\rangle 
\langle j,n+1|) ]. \label{ep}
\en
In this procedure, the two noninteracting single-mode leads of Eq.~(\ref{lead}) are 
mapped to a multichannel model
\bq
\widetilde{H}_{\rm lead}=\sum_{\eta, k, n} \epsilon_{\eta k n} |\eta k, n\rangle \langle 
\eta k, n|
\eq
with $\epsilon_{\eta k n}=\epsilon_{\eta k}+ n \hbar \omega_{ph}$. Here, the difference 
from traditional multi-channel leads lies in the fact that the channel index $n$ 
represents the phonon quanta excited in the device, thus generating a weight factor 
$P_n=(1-e^{-\beta \omega_{ph}})e^{-n\beta \omega_{ph}}$ [the statistical probability of 
the phonon number state $|n\rangle$ at finite temperature $T$ ($\beta=1/k_{\rm B}T$)], 
which must be associated with this channel. In this connection, we assume that the 
phonon bath is perpetually in thermal equilibrium, even under nonequilibrium transport 
conditions, due to a high energy relaxation rate. Thus, we ignore nonequilibrium phonon 
effects, considering phonon equilibration fast compared to the dwell time of an electron 
on the QD.\cite{Mitra} We also note that different channels are admixed with each other 
due to the EPI term, Eq.~(\ref{ep}), which takes place solely in the central region. 
Finally, the tunneling part (\ref{tunneling}) can also be rewritten in terms of this 
basis set:
\bq
\widetilde{H}_{\rm tl}=\sum_{\eta,k,j,n} V_{\eta, j}^{n}(|\eta k,n\rangle \langle 
j,n|+{\rm H.c.}).
\eq
$V_{\eta,j}^{n}$ is the coupling between the $n$th pseudochannel in lead $\eta$ and the 
device. Clearly, the many-body EPI problem is transformed to a multichannel 
single-electron scattering problem with the help of the new representation, as 
illustrated in Fig.~1.

\begin{figure}[htb]
\includegraphics [width=7cm,height=10cm,angle=0,clip=on] {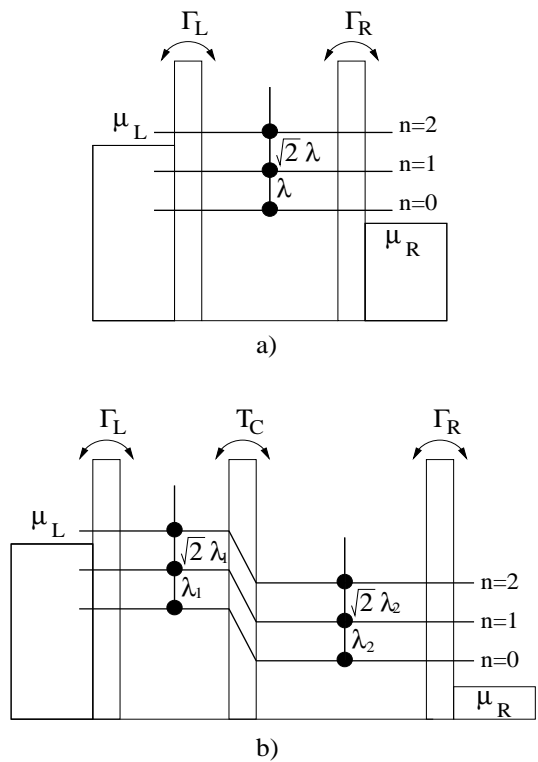}
\caption{Schematic description of the inelastic scattering problem for (a) a single site  
and (b) two sites in series, both with on-site EPI. Each phonon state of the site, taken 
jointly with the Bloch state of the electron in the lead, can be visualized as a 
pseudochannel labeled by $n$, which is connected to the two leads with hopping 
parameters $\Gamma_{L/R}$. These channels are connected vertically by the EPI, 
$\lambda_{j}$.} \label{fig1}
\end{figure}

We depict the physical process of tunneling after remodeling the electron-phonon 
coupling system as follows: The $n$th pseudochannel in either the left or the right lead 
denotes an electron propagating in one of the single-electron modes of that lead with 
the device being in its $n$-phonon state. An electron incident on pseudochannel $n$ in 
the left or right lead can transport elastically to the central region, where it then 
excites or absorbs $m$ phonons and exits inelastically into the $(n\pm m)$th channel in 
the left or right lead; or experiences no exchange with phonon and exits elastically in 
the same channel in both leads. It is of central importance to note that, within this 
formulation, the many-body problem can be solved exactly and consequently there is no 
loss of phase coherence in the treatment of the electron-phonon exchange. Indeed, such 
coherent interference effects in the emission processes of phonons play a crucial role 
in nonlinear transport in CQDs. This is the reason that the present approach can provide 
a satisfactory description of the current spectrum consistent with actual experimental 
results.\cite{Fujisawa}

\subsection{Current-current fluctuation spectra} 

In the section, we derive a general formula describing low-frequency current fluctuation 
spectra for a two-terminal mesoscopic device coupled to an Einstein-phonon bath (with 
frequency $\omega_{ph}$). In the combined electron-phonon representation, the many-body 
problem becomes a one-body multichannel scattering problem, as described in the last 
section. With this picture in mind, we employ the B\"uttiker scattering approach to 
achieve this objective.\cite{Buttiker,Blanter}   

First we introduce creation and annihilation operators of electrons for the incoming and 
outgoing states in all transverse channels. We employ creation, $a_{\eta 
n}^{\dagger}(\omega)$, and annihilation, $a_{\eta n}(\omega)$, operators for electrons 
with total energy $\omega$ incident upon the central region from channel $n$ of lead 
$\eta$; and we employ the creation, $b_{\eta n}^{\dagger}(\omega)$, and annihilation, 
$b_{\eta n}(\omega)$, operators for electrons in the outgoing states. Clearly, they obey 
anticommutation relations.

In second quantized notation, we can write explicit forms for the wave function 
$\Psi_{\eta}({\bf r},t)$ in lead $\eta$ as:
\bn
\Psi_{\eta}({\bf r},t)&=& \int d\omega e^{-i \omega t/\hbar} \sum_{n=1}^N 
\frac{\chi_{\eta n}({\bf r}_{\perp})}{(2\pi \hbar v_{\eta n}(\omega))^{1/2}} \cr
&& \times [a_{\eta n}(\omega) e^{ik_{\eta n}z} + b_{\eta n}(\omega) e^{-ik_{\eta n}z}]. 
\label{wf}
\en
Here, $\chi_{\eta n}$ are the transverse parts of the electron wave functions [${\bf 
r}=({\bf r}_{\perp},z)$], the wave vector in the $n$th channel of lead $\eta$ is 
$k_{\eta n}=\hbar^{-1}[2m^* (\omega - n\hbar \omega_{ph})]^{1/2}$ ($m^*$ is the 
effective electron mass), and the corresponding electron velocity is $v_{\eta 
n}(\omega)=\hbar k_{\eta n}/m^*$. $N$ denotes the total phonon number, which is infinite 
in principle, but must truncated to a finite value for computational purposes. 
Correspondingly, we define the current operator as:
\bn
{\hat I}_{\eta}(z,t)&=&\frac{\hbar e}{2im} \int d{\bf r}_{\perp} \left [ 
\Psi_{\eta}^{\dagger}({\bf r},t) \frac{\partial}{\partial z} \Psi_{\eta}({\bf r},t) 
\right. \cr
&& \left. - \left ( \frac{\partial}{\partial z} \Psi_{\eta}^{\dagger}({\bf r},t) \right 
) \Psi_{\eta}({\bf r},t) \right ].
\en
Substituting the wave function of Eq.~(\ref{wf}) into this definition, and assuming that 
the velocities $v_{\eta n}(\omega)$ vary with energy quite slowly, the current operator 
can be simplified as:
\bn
{\hat I}_{\eta}(t)&=&\frac{e}{h} \sum_{n} \int d\omega d\omega' 
e^{i(\omega-\omega')t/\hbar} [a_{\eta n}^{\dagger}(\omega) a_{\eta n}^{\pdag}(\omega') 
\cr
&& - b_{\eta n}^{\dagger}(\omega) b_{\eta n}^{\pdag}(\omega')].
\en
In the scattering approach, the outgoing electron operators, $b$, are related to the 
incoming operators, $a$, via the $2N\times 2N$ scattering matrix ${\bm s}$:
\bq
{\bm s}=\left (
\begin{array}{cc}
{\bm s}_{L,L} & {\bm s}_{L,R} \\
{\bm s}_{R,L} & {\bm s}_{R,R}
\end{array}
\right ) = \left (
\begin{array}{cc}
{\bm r}_L & {\bm t}_L \\
{\bm t}_R & {\bm r}_R
\end{array}
\right ),
\eq
in which all blocks have $N\times N$ dimensions. An element $s_{\eta n,\eta m}$ of the 
diagonal blocks describes a tunneling event in which an electron is incident from the 
$\eta m$ channel and reflects back to the $n$th channel in the same lead. For this 
reason we denote it as $r_{Lnm}$ ($r_{Rnm}$), the reflection amplitude for the left 
(right) lead. Similarly, an element $s_{\eta n,\eta' m}$ ($\eta' \neq \eta$) of the 
off-diagonal blocks denotes electron transmission through the central region from the 
$\eta' m$ channel to the $\eta n$ channel; and these are the transmission amplitudes 
$t_{Lnm}$ and $t_{Rnm}$. It should be noted that the scattering matrix is unitary ${\bm 
s}^{\dagger} {\bm s}={\bm I}$ (${\bm I}$ is the $N$-dimensional unit matrix) and 
symmetric. Therefore, we can express the current in terms of the operators $a$ and 
$a^{\dagger}$ alone: 
\bn
{\hat I}_{\eta}(t)&=& \frac{e}{h} \sum_{\alpha \beta, m n} \int d\omega d\omega' 
e^{\frac{i}{\hbar}(\omega-\omega')t} a_{\alpha m}^{\dagger}(\omega) \cr
&& \times A_{\alpha \beta}^{m n}(\eta; \omega,\omega') a_{\beta n}^{\pdag}(\omega'),
\en
where
\bq
A_{\alpha \beta}^{m n}(\eta; \omega,\omega')=\delta_{mn} \delta_{\alpha \eta} 
\delta_{\beta \eta} - \sum_{l} s_{\eta m, \alpha l}^{\dagger}(\omega) s_{\eta l, \beta 
n}^{\pdag}(\omega'). \label{a}
\eq

Proceeding to evaluation of the shot noise spectra, we recall that the noise power 
spectra between the leads $\eta$ and $\eta'$ can be expressed as the Fourier transform 
of the current-current correlation function: 
\bn
\label{powerspectrum}
S_{\eta \eta'}(\omega) &=& \frac{1}{2}\int_{-\infty}^{\infty} dt e^{i \omega (t-t')} 
[\langle {\hat I}_{\eta}(t) {\hat I}_{\eta'}(t') \rangle \cr
&& + \langle
{\hat I}_{\eta'}(t') {\hat I}_{\eta}(t) \rangle - 2 \langle {\hat I}_{\eta}(t) \rangle 
\langle {\hat I}_{\eta'}(t') \rangle ] , 
\en
where $\langle \cdots \rangle$ represents the quantum statistical average. Focusing 
attention on low-frequency noise in the same lead in the present paper, we obtain, after 
some algebra\cite{Buttiker,Blanter}
\bn
S_{\eta \eta}(0) &=& \frac{e^2}{h} \sum_{\alpha\beta, mn} \sum_{\mu\nu, kl} \int d\omega 
d\omega' A_{\alpha\beta}^{mn}(\eta; \omega, \omega) \cr
&& \hspace{-0.8cm}\times A_{\mu\nu}^{kl}(\eta;\omega',\omega') \left \{ \langle 
a_{\alpha m}^{\dagger}(\omega) a_{\nu l}^{\pdag}(\omega')\rangle \langle a_{\beta 
n}^{\pdag}(\omega) a_{\mu k}^{\dagger}(\omega')\rangle \right. \cr
&& \hspace{-0.8cm} \left. + \langle a_{\mu k}^{\dagger}(\omega') a_{\beta 
n}^{\pdag}(\omega) \rangle \langle a_{\nu l}^{\pdag}(\omega') a_{\alpha 
m}^{\dagger}(\omega)\rangle \right \}.
\en
Considering the fact that the channel actually corresponds to the phonon number state in 
this mapping model, these quantum statistical averages, $\langle a^{\dagger} a \rangle$, 
must be calculated with caution: an additional weight factor $P_n$ must accompany any 
starting channel $n$ jointly with the Fermi distribution, $f_{\eta}^n(\omega)=[ 1+ 
e^{(\omega+ n\hbar \omega_{ph}-\mu_{\eta})/k_{\rm B}T}]^{-1}$. Therefore, we have 
\bn
S_{\eta \eta}(0) &=& \frac{e^2}{h} \sum_{\alpha\beta, mn} \int d\omega 
A_{\alpha\beta}^{mn}(\eta;\omega,\omega) A_{\beta\alpha}^{nm}(\eta;\omega,\omega) \cr
&& \hspace{-1.8cm} \times \left \{ P_{m}f_{\alpha}^{m}(\omega) [1- 
f_{\beta}^{n}(\omega)] + P_n f_{\beta}^{n}(\omega) [1- f_{\alpha}^{m}(\omega)] \right 
\}. \label{sn}
\en
In contrast to the case without EPI, one can not further simplify the shot noise formula 
of Eq.~(\ref{sn}) to write it in terms of transmission probabilities only without any 
detailed information concerning the device. Here, the shot noise formula for inelastic 
tunneling requires information about the transmission and reflection amplitudes of all 
pseudochannels. Actually, a similar result, that the noise can not be expressed in terms 
of the transmission probabilities only in the boson-field assisted tunneling, has been 
addressed in the studies of ac-driven current noise in quantum dots.\cite{Camalet} 
In the next two sections, we will apply this theory to evaluate zero-frequency shot 
noise in a single-molecular QD and semiconductor CQDs based on Eqs.~(\ref{a}) and 
(\ref{sn}).      

It should be noted that since the mapping technique involves an underlying 
single-particle picture for the dot EPI, it does ignore the possibility of strong 
bath-dot many-body effects which could invalidate the single-particle picture, 
introducing broadening of tunneling associated with strong EPI and nonequilibrated 
phonon.\cite{Flensberg,Mitra} Such many-body effects are understood to play an important 
role in tunneling in the linear regime, but become trivial in the case of high 
bias-voltages, upon which this paper is focused.

\section{A single-molecular quantum dot}

\subsection{Model and shot noise formula}

In this section, we examine the EPI effects on the zero-frequency noise properties of a 
single-molecular QD. For this single-site case [shown in Fig.~1(a)], the Hamiltonian 
$H_{d}$ of the central region becomes
\bq
H_{d}=\epsilon_{d} d_1^\dagger d_1^{\pdag},
\eq
and summation with respect to the site index $j$ reduces to only one term. In terms of 
the combined electron-phonon representation, we rewrite it as:
\bq
\widetilde{H}_{d}=\sum_{n\geq 0} (\epsilon_{d}+ n\hbar \omega_{ph}) |1,n\rangle \langle 
1,n|. 
\eq

As suggested by Haule and Bon\u ca in their original work,\cite{Bonca} the transmission 
and reflection amplitudes in this single-site case can be obtained by solving the 
Schr\"odinger equation with trial wave functions. In our recent study of low-temperature 
time-dependent phonon-assisted tunneling through a single-molecular QD, we adopted the 
NGF method to compute the time-depdendent transmission probabilities within the 
framework of the mapping technique.\cite{Dongac} In fact, Fisher and Lee established a 
general relation between the scattering matrix elements and the retarded GFs of the 
conductor.\cite{Fisher} Hence, in the present paper we will employ the NGF technique to 
compute these transmission and reflection amplitudes in the wide band limit. 

Following our previous paper,\cite{Dongac} we define pseudo-Fermi operators:
\bq
d_{1n}^{\dagger}=|1,n\rangle \langle 0|, \\\\ c_{\eta k n}^{\dagger}=|\eta k, n\rangle 
\langle 0|.
\eq
and rewrite the mapping Hamiltonian in terms of these pseudoperators. The evaluation of 
the retarded GFs of the QD, $G_{mn}^{r}(t,t')\equiv i \theta(t-t')\langle \{d_{1m}(t), \ 
d_{1n}^{\dagger}(t') \}\rangle$, proceeds using the equation-of-motion technique. In the 
Fourier space, we have
\bn
&& [\omega - \epsilon_d -m\hbar \omega_{ph} -\Sigma_m^r(\omega)] G_{mn}^r(\omega) = 
\delta_{mn} \cr
&& - \lambda_1 \sqrt{m} G_{(m-1)n}^r(\omega) - \lambda_1 \sqrt{m+1} 
G_{(m+1)n}^r(\omega), \label{gfrome}
\en 
where the retarded self-energy $\Sigma_{m}^r(\omega)$ due to coupling to the leads is 
defined as
\bq
\Sigma_{m}^{r}(\omega)=\sum_{\eta,k}\frac{|V_{\eta m}|^2} {\omega-\epsilon_{\eta k} - m 
\hbar \omega_{ph}+ i 0^{+}}.
\eq
In the wide band approximation, where the hopping matrix element $V_{\eta j}^n = V_{\eta 
j}$ is independent of energy, one has $\Sigma_{m}^{r} = - \frac{i}{2} (\Gamma_{L} + 
\Gamma_{R})$ with $\Gamma_{\eta}=2\pi \sum_{k} |V_{\eta}|^2 \delta(\omega-\epsilon_{\eta 
k}-m \hbar \omega_{ph})$ as the generalized linewidth function. For simplicity, we 
assume the tunnel couplings between the molecular QD and the two leads to be symmetric 
$\Gamma_{L}=\Gamma_{R}=\Gamma$. Therefore, one can rewrite the resulting 
Eq.~(\ref{gfrome}) in a compact matrix form:
\bq
[(\omega+i\Gamma){\bm I}-{\bm B}] {\bm G}^{r}(\omega)={\bm I}, \label{gfrsqd}
\eq
in which ${\bm B}$ is a $N\times N$ symmetrical tridiagonal matrix with 
$B_{nn}=\epsilon_{d}+n\hbar \omega_{ph}$, $B_{n(n-1)}=-\lambda \sqrt{n}$, and 
$B_{n(n+1)}=-\lambda \sqrt{n+1}$. A similar result is obtained for the advanced GF,  
${\bm G}^a(\omega)$, and we have ${\bm G}^a=[{\bm G}^r]^{\dagger}$ in $\omega$-space. 
The element $G_{nn}$ corresponds to the probability amplitude of propagating in a state 
that starts and ends with the same number of $n$ phonons, while an element $G_{mn}$ 
($n\neq m$) denotes the probability amplitude of starting with $m$ and ending with $n$ 
phonons.

Furthermore, employing the Fisher-Lee relation:\cite{Fisher}
\bq
{\bm s}(\omega)=-{\bm I}+ i \sqrt{\Gamma_{L} \Gamma_{R}} {\bm G}^{r}(\omega), 
\label{flr}
\eq  
we immediately obtain the transmission and reflection amplitudes in the presence of a 
phonon bath:
\bn
&&s_{L(R)m,L(R)n}=r_{L(R)mn}=r_{mn}=-\delta_{mn}+i\Gamma G_{mn}^r(\omega), \cr
&&s_{L(R)m,R(L)n}=t_{R(L)mn}=t_{mn}=i\Gamma G_{mn}^r(\omega). \label{tra}
\en
It is easy to check that they obey properties (1) ${\bm r}^{\dagger} {\bm r} + {\bm 
t}^{\dagger} {\bm t} ={\bm I}$; (2) ${\bm r}^{\dagger} {\bm t} + {\bm t}^{\dagger} {\bm 
r} =0$, due to the unitary property of the scattering matrix; and (3) ${\bm r}=-{\bm 
I}+{\bm t}$; (4) $2 {\bm t}^{\dagger}{\bm t} = {\bm t}^{\dagger} + {\bm t}$.     

Inserting the resulting transmission and reflection amplitudes, Eq.~(\ref{tra}), into 
Eqs.~(\ref{a}) and (\ref{sn}), we obtain, after some algebra:
\begin{subequations}
\label{snsqd}
\bn
&&S_{LL}(0)=S_{RR}(0)= \frac{2e^2 }{h} \int d\omega \sum_{mn} \left \{ [({\bm 
t}^{\dagger} {\bm t})_{mn}]^2 \right. \cr
&& \times \left [P_n f_{L}^n(\omega) (1- f_{L}^m(\omega)) + P_m f_R^m(\omega)(1- 
f_{R}^n(\omega)) \right ] \cr
&& + ({\bm r}^{\dagger} {\bm r})_{mn} ({\bm t}^{\dagger} {\bm t})_{mn} [P_n 
f_{L}^n(\omega)(1- f_{R}^m(\omega)) \cr
&&\left. + P_m f_R^m(\omega) (1- f_L^n(\omega))] \right\},
\en
and $S_{LR}(0)=S_{RL}(0)=-S_{LL}(0)$. We can also determine the elastic contribution to 
shot noise by imposing the constraint of elastic tunneling $m=n$:
\bn
&&S_{LL}^{\rm el}(0)= \frac{2e^2 }{h} \int d\omega \sum_{n} \left \{ ({\bm t}^{\dagger} 
{\bm t})_{nn} \right. \cr
&& \times  P_n [f_{L}^n(\omega) (1- f_{R}^n(\omega)) + f_R^n(\omega)(1- 
f_{L}^n(\omega))] \cr
&& \left. - [({\bm t}^{\dagger} {\bm t})_{nn}]^2 P_n [ f_{L}^n(\omega)- 
f_{R}^n(\omega)]^2 
\right\}. 
\en
\end{subequations}
Thus, the total current $I_{L}$ and its elastic part are given by:\cite{Bonca2,Dongac}
\begin{subequations}
\bn
I_{L}&=&\frac{e}{h} \int d\omega \sum_{mn} ({\bm t}^{\dagger} {\bm t})_{mn} 
\hspace{2cm}\cr
&& \hspace{-1.5cm}\times \left \{ P_n f_{L}^n(\omega) [1-f_{R}^m(\omega)] - P_m 
f_{R}^m(\omega) [1-f_{L}^n(\omega)] \right \},
\en   
and
\bq
I_{L}^{\rm el}=\frac{e}{h} \int d\omega \sum_{n} ({\bm t}^{\dagger} {\bm t})_{mn} P_n 
[f_{L}^n(\omega) - f_{R}^n(\omega)].
\eq
\end{subequations}
Furthermore, the Fano factor, which measures the deviation from uncorrelated Poissonian 
noise, is defined as:
\bq
F=\frac{S_{LL}(0)}{2eI_{L}}.
\eq

\subsection{Numerical results and discussions}

In this subsection we address the numerical calculation of zero-frequency shot noise 
through a single-molecular QD in the presence of a phonon bath, on the basis of 
Eqs.~(\ref{gfrsqd}), (\ref{tra}) and (\ref{snsqd}). For simplicity, we assume the bias 
voltage is distributed symmetrically, i.e., $+V/2$ on the left lead and $-V/2$ on the 
right lead. We also set the phonon energy $\omega_{ph}=1$ as the unit of energy 
throughout the rest of the paper and choose the Fermi levels of the two leads as the 
reference of energy $\mu_{L}=\mu_{R}=0$ at equilibrium. Other parameters in our 
calculations for the single-molecular QD are: $\lambda=0.5$, $\Gamma=0.04$, and 
$T=0.04$.

It should be emphasized that the present approach is an {\em exact} method to deal with 
an interacting electron-phonon system having arbitrary coupling strength if the maximum 
number of phonons $N_{ph}$ involved in the calculation is taken up to infinity. 
Unfortunately this is impossible in a real calculation and we have to truncate this 
number to a finite value chosen to insure computational convergence with desired 
accuracy. The appropriate value of $N_{ph}$ depends on the energy of the Einstein-phonon 
mode, the EPI constant, and the temperature of the system under investigation. For the 
parameters involved in the present paper, we choose $N_{ph}=8$ to obtain results with 
accuracy better than $1\%$ in this and next sections.     

\begin{figure}[htb]
\includegraphics [width=7cm,height=14cm,angle=0,clip=on] {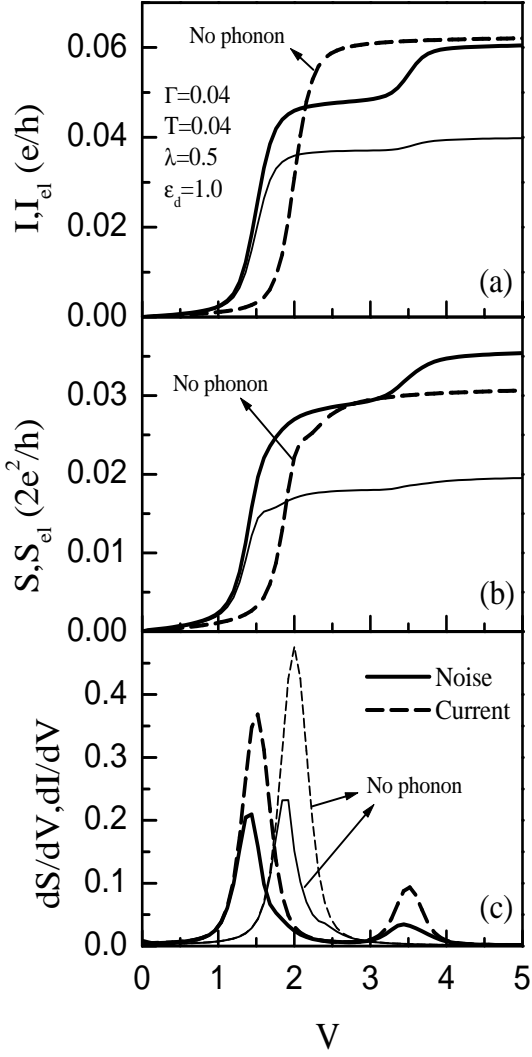}
\caption{(a) The calculated total current $I$ (thick line), elastic current $I^{\rm el}$ 
(thin line); (b) the zero-frequency shot noise $S$ (thick line), its elastic part 
$S^{\rm el}$ (thin line); and (c) the corresponding differential conductance $dI/dV$ 
(dashed line) and differential shot noise $dS/dV$ (solid line) as functions of applied 
bias voltage for a single-molecular QD with bare level $\epsilon_{d}=1.0$. For 
comparison, we also plot the current and shot noise in the case without the phonon bath. 
The other parameters used in calculation are: $\Gamma=0.04$, $\omega_{ph}=1.0$, and 
$\lambda=0.5$. The temperature is set to $T=0.04$.} \label{fig2}
\end{figure}

In Fig.\,2 we plot the total and elastic currents (a) and shot noises (b) as functions 
of the bias voltage $V$. For comparison, we also plot the current and shot noise for the 
same system without the EPI as dashed lines. Obviously, the shot noise and current have 
similar step characteristics with increasing bias-voltage. When the electron is not 
coupled to the phonon mode, only one step structure occurs when the bias-voltage matches 
the Fermi energy of the left lead with the single level in the QD, leading to the single 
peak in the differential conductance $dI/dV$ and the differential shot noise $dS/dV$ 
shown in Fig.~2(c). In the presence of a phonon bath, we find that: (1) The overall 
spectra are both shifted by $\lambda^2/\omega_{ph}$ due to the polaron effect (for 
instance, the main peaks in $dI/dV$ vs. $V$ and $dS/dV$ vs. $V$ corresponding to 
resonance with the QD level shift from the position $V/2=\epsilon_{d}$ to 
$\epsilon_{d}-\lambda^2/\omega_{ph}$, with slightly suppressed amplitudes); and (2) new 
resonant peaks separated by the frequency of the phonon mode, $\hbar \omega_{ph}$, 
appear due to emission of phonons, indicating that a new pseudochannel has opened and 
participates in contributing to tunneling.    

\begin{figure}[htb]
\includegraphics [width=5.cm,height=5cm,angle=0,clip=on] {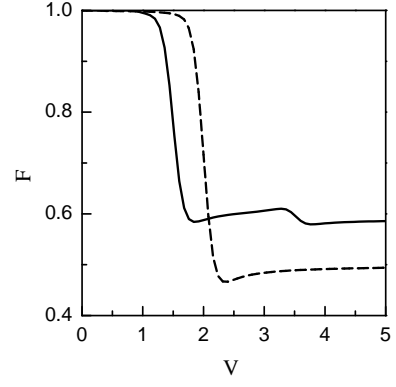}
\caption{Fano factor vs. bias-voltage $V$. Solid line denotes results with EPI, while 
the dashed line denotes results without EPI. The parameters are the same as in Fig.\,2.} 
\label{fig3}
\end{figure}

Figure 3 shows the bias-dependent Fano factor for the system under study. Roughly 
speaking, it is known that the current fluctuation for a noninteracting two-terminal 
conductor is $\sum_{n} T_{n} (1-T_{n})$ at zero temperature, and the corresponding Fano 
factor $F$ is proportional to $\sum_{n} T_{n} (1-T_{n})/\sum_{n} T_{n}$, where $T_{n}$ 
is the transmission probability of channel $n$.\cite{Buttiker,Blanter} Therefore, there 
is no current noise if the transmission probability is $T_n=0$ or $1$, and the Fano 
factor is correspondingly equal to $F=1$ or $0$, respectively. This is case in the low 
bias-voltage region, below the single-level resonance. In contrast, in the high 
bias-voltage limit, the Fano factor tends to $\frac{1}{2}$ for a symmetric 
noninteracting system, $\Gamma_{L}=\Gamma_{R}$. In addition, we observe that the Fano 
factor is significantly enhanced and exhibits more fine structure due to phonon effects 
in this region, in comparison with the uncorrelated noise.

\section{Double quantum dots}

Finally, we examine nonlinear transport and fluctuations in semiconductor CQDs in a 
series arrangement, for which $V_{L2}^n=V_{R1}^n=0$. As shown in Fig.~1(b), the 
Hamiltonian $H_{d}$ for the two-site case is given by:
\bq
H_{d}=\sum_{j=1,2}\epsilon_{j} d_j^\dagger d_j^{\pdag} - T_c (d_{1}^{\dagger} d_{2} + 
{\rm H.c.}),
\eq
where $T_{c}$ denotes the tunnel coupling between the two QDs. In terms of the joint 
electron-phonon representation, it becomes:
\bn
\widetilde{H}_{d}&=&\sum_{j,n\geq 0} (\epsilon_{j}+ n\hbar \omega_{ph}) |j,n\rangle 
\langle j,n| \cr
&& - \sum_{n\geq 0} T_c (|1,n\rangle \langle 2,n| + {\rm H.c.}). 
\en
We also ignore higher-order tunneling processes by assuming that coherent electron 
transfer between the two QDs, governed by $T_{c}$, does not explicitly involve 
electron-phonon exchange. Correspondingly, the retarded GFs can be written in a 
block-matrix form:
\bn
&& \left (
\begin{array}{cc}
{\bm G}_{11}^r(\omega) & {\bm G}_{12}^r(\omega) \\
{\bm G}_{21}^r(\omega) & {\bm G}_{22}^r(\omega)
\end{array}
\right ) = \cr
&& \left (
\begin{array}{cc}
(\omega+i\Gamma_{L}){\bm I}-{\bm B}_{1} & {\bm C} \\
{\bm C} &  (\omega+i\Gamma_{R}){\bm I}-{\bm B}_{2}
\end{array}
\right )^{-1},
\label{gfrcqd}
\en
in which ${\bm B}_{j}$ ($j=1,2$) is a $N\times N$ symmetrical tridiagonal matrix: 
$B_{j,nn}=\epsilon_{j}+n\hbar \omega_{ph}$, $B_{j,n(n-1)}=-\lambda_{j} \sqrt{n}$, and 
$B_{j,n(n+1)}=-\lambda_{j} \sqrt{n+1}$, and ${\bm C}$ is a $N$-dimensional diagonal 
matrix $C_{mn}=-\delta_{mn}T_{c}$. Once again, we can write the transmission and 
reflection amplitudes through the CQDs in terms of the retarded GFs, ${\bm 
G}_{ij}^r(\omega)$:
\bn
&&s_{L(R)m,L(R)n}=r_{L(R)mn}=-\delta_{mn}+i\Gamma G_{11(22),mn}^r(\omega), \cr
&&s_{L(R)m,R(L)n}=t_{R(L)mn}=i\Gamma G_{12(21),mn}^r(\omega). \label{tracqd}
\en
Considering the properties of the matrices involved, we simplify the total current 
$I_{L}$ as:
\bn
I_{L}&=&\frac{e}{h} \int d\omega \sum_{mn} \left \{ ({\bm t}_L^{\dagger} {\bm t}_L)_{mn} 
P_n f_{L}^n(\omega) (1-f_{R}^m(\omega)) \right. \cr
&& \left. - ({\bm t}_R^{\dagger} {\bm t}_R)_{nm} P_m f_{R}^m(\omega) (1-f_{L}^n(\omega)) 
\right \} \cr
&=& \frac{e}{h} \int d\omega \sum_{mn} ({\bm t}_L^{\dagger} {\bm t}_L)_{mn} \left \{ P_n 
f_{L}^n(\omega) (1-f_{R}^m(\omega)) \right. \cr
&& \left. - P_m f_{R}^m(\omega) (1-f_{L}^n(\omega)) \right \}. \label{curcqd}
\en   
Furthermore, the zero-frequency shot noise $S_{LL}(0)$ is given by
\bn
&&S_{LL}(0)= \frac{2e^2 }{h} \int d\omega \sum_{mn} \left \{ |({\bm t}_{L}^{\dagger} 
{\bm t}_{L})_{mn}|^2 \right. [P_n f_{L}^n(\omega) \cr
&& \times (1- f_{L}^m(\omega)) + P_m f_R^m(\omega)(1- f_{R}^n(\omega))] \cr
&& + |({\bm t}_{R}^{\dagger} {\bm r}_{L})_{mn}|^2 [P_n f_{L}^n(\omega)(1- 
f_{R}^m(\omega)) \cr
&&\left. + P_m f_R^m(\omega) (1- f_L^n(\omega))] \right\}, \label{sncqd}
\en
and also $S_{LR}(0)=S_{RL}(0)=-S_{LL}(0)=-S_{RR}(0)$. Their elastic contributions can be 
obtained by setting $m=n$. 

To compare this theoretical analysis with experimental results, we numerically calculate 
the current and shot noise based on Eqs.~(\ref{curcqd}) and (\ref{sncqd}) as functions 
of the energy difference between the two QDs, $\varepsilon=\epsilon_{1}-\epsilon_{2}$, 
which can be tuned by varying the gate voltage. We address only the case of high 
bias-voltage, $V=\mu_{L}-\mu_{R} \gg 0$. While the present formalism is valid for 
arbitrary choice of coupling parameters, tunneling strengths $\Gamma$ and $T_{c}$, as 
well as EPI constants $\lambda_{1(2)}$, we specialize in the interest of comparison with 
the CQD experiments,\cite{Fujisawa} choosing these parameters as: 
$\Gamma_{L}=\Gamma_{R}=\Gamma=0.25$, $T_{c}=0.25$, 
$\lambda_1=-\lambda_2=\lambda=0.1875$, and temperature $T=0.125$ ($\omega_{ph}=1$). In 
our calculations we set $eV=10$ so that $\mu_{L} \gg \epsilon_1, \epsilon_2 \gg 
\mu_{R}$, in which case nonlinear transport does not significantly depend on $\mu_{L}$ 
and $\mu_{R}$ at low temperatures.  

In Fig.~4(a), we plot the resulting currents, including the total current, its elastic 
and inelastic components, and the current without EPI, as functions of energy 
difference, $\varepsilon$. The external voltage, $V$, gives rise to a finite tunneling 
current through the CQDs. In the case without EPI, one can observe a single maximum in 
the current spectrum at $\varepsilon=0$, i.e., the resonance point at which the energy 
level of the left dot is matched with that of the right dot, and there is a rapid 
decrease upon departure from this point. In contrast, the inelastic current due to 
phonon coupling is remarkably enhanced on the positive branch of the energy difference, 
$\varepsilon> 0$, in which phonons are emitted during the tunneling process. One can 
even unambiguously observe some peaks when the condition for spontaneous 
phonon-emission-assisted resonance, $\varepsilon=n \hbar \omega_{ph}$ ($n=1,2$ in our 
figure), is satisfied. However, for negative energy gap, $\varepsilon< 0$, the 
contribution of phonon absorption processes to inelastic current is negligible because 
there are not enough phonons available to be absorbed at low temperature. Moreover, our 
calculations predict a shoulderlike structure for $\varepsilon< 0$, which arises solely 
from elastic channels in the presence of the phonon bath.       

\begin{figure}[htb]
\includegraphics [width=8.5cm,height=6.5cm,angle=0,clip=on] {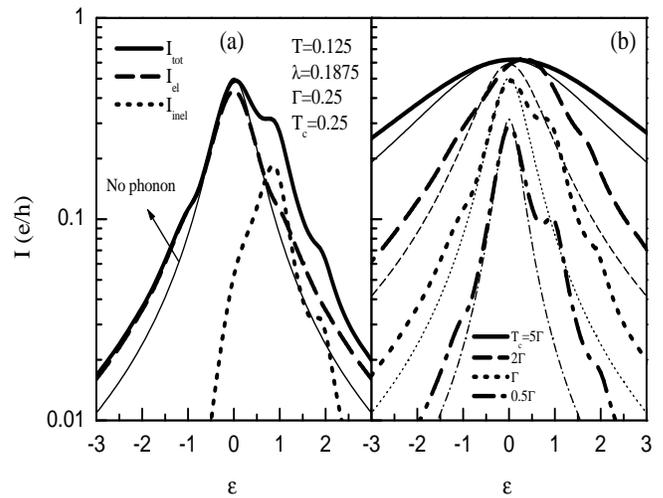}
\caption{Predicted tunneling current as a function of the energy difference, 
$\varepsilon$. (a) The total current $I_{\rm tot}$, its elastic component $I_{\rm el}$, 
its inelastic part $I_{\rm inel}$, and the current without phonon bath (thin line) vs. 
$\varepsilon$ (with $\Gamma=T_{c}=0.25$, EPI constants $\lambda_1=-\lambda_2=0.1875$, 
and temperature $T=0.125$); (b) The total currents (thick lines) and currents without 
EPI (thin lines) for various couplings, $T_{c}$, between the two QDs.} \label{fig4}
\end{figure}

It should be mentioned that the asymmetric and oscillatory shoulder behaviors in the 
current spectrum have also been obtained by means of the master equation\cite{Brandes} 
and the nonperturbative real-time renormalization-group (RTRG) method.\cite{Keil} 
Nevertheless, the assumption taken in the master equation method (i.e., weak tunnel 
coupling) limits its validity for analysis of the current spectrum at stronger values of 
interdot coupling, $T_{c}$. Albeit the RTRG approach has not such limitation, this 
problem was not actually studied in Ref.~[\onlinecite{Keil}]. Once again, we emphasize 
that our studies also assume the phonon bath to be in thermal equilibrium perpetually. 
Treatment of nonequilibrium phonon effects for CQDs is beyond the scope of the present 
paper. However, one of the advantages of the present scheme is that it treats the 
internal tunneling, $T_{c}$, on an equal footing with the couplings to the leads as well 
as EPI. Thus, it allows us to calculate the current spectrum for arbitrary internal 
tunneling strength, $T_{c}$. Such evaluations are exhibited in Fig.~4(b). With 
increasing $T_{c}$, the inelastic-resonance-induced oscillating shoulder structure is 
gradually smeared out, in agreement with experimental observations.\cite{Fujisawa}

\begin{figure}[htb]
\includegraphics [width=8.5cm,height=6.5cm,angle=0,clip=on] {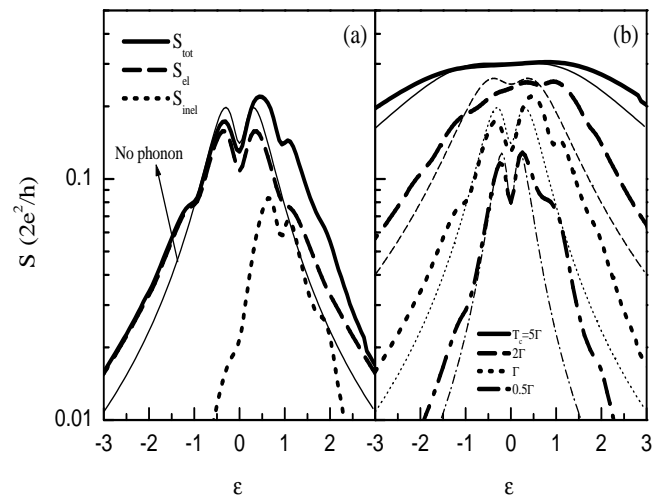}
\caption{Zero-frequency shot noise vs. $\varepsilon$. (a) Total shot noise, $S_{\rm 
tot}$; its elastic part, $S_{\rm el}$; its inelastic part, $S_{\rm inel}$; and shot 
noise with no EPI (thin line). The parameters are the same as in Fig.~4(a); (b) Shot 
noise with (thick lines) and without (thin lines) EPI for several different values of 
$T_{c}$.} \label{fig5}
\end{figure}

In regard to zero-frequency shot noise, our evaluations are summarized in Fig.~5. In 
Fig.~5(a), where we plot the calculated shot noise spectra for the same system studied 
in Fig.~4(a), we see that the shot noise without EPI (thin curve) exhibits two peaks 
located symmetrically around the resonant point, $\varepsilon=0$. This behavior is due 
to the fact that there is no noise generation when the transmission probability is 
$T_{tr}=0$ or $1$, while maximal generation occurs between these values. In contrast to 
this, we find that phonon coupling enhances the shot noise peak located at the positive 
$\varepsilon$-side ($\varepsilon> 0$), but reduces the peak on the negative side, 
leading to breakdown of the symmetric structure. Moreover, interference of the EPI in 
the CQDs gives rise to oscillatory shoulders. It should be noted that, since the 
transmission probability corresponding to these inelastic resonances is small, the shot 
noise spectrum exhibits only a single peak rather than two peaks in the phonon-assisted 
resonant tunneling region.   

\begin{figure}[htb]
\includegraphics [width=8.5cm,height=6cm,angle=0,clip=on] {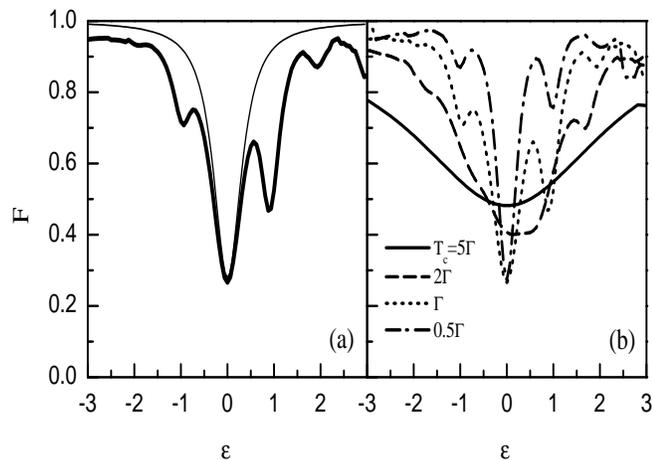}
\caption{Fano factor vs. $\varepsilon$. (a) Fano factors with (thick curve) and without 
(thin curve) EPI for the same system as in Fig.~4(a); (b) Fano factors for several 
different values of $T_{c}$.} \label{fig6}
\end{figure}

The resulting Fano factor is plotted as a function of the energy gap, $\varepsilon$, in 
Fig.~6. At the elastic resonance point, $\varepsilon=0$, the smallest Fano factor is 
observed, implying that quantum coherence significantly suppresses shot noise. For 
inelastic resonant tunneling, spontaneous phonon emission processes also reduce shot 
noise, which induces additional dips in the Fano factor for $\varepsilon>0$. 
Interestingly, our results predict fine structure in the negative branch of 
$\varepsilon$. This is due to the shoulders in the current and shot noise spectra 
stemming from the contributions of elastic channels. It indicates that the Fano factor 
is quite sensitive to the interaction with phonons. Moreover, for large $|\varepsilon|$, 
the shot noise is dominated by the Poisson process, i.e., Fano factor $F\rightarrow 1$. 
Finally, increasing the interdot tunneling, $T_{c}$, progressively smooths out the shot 
noise spectra and the Fano factors as displayed in Figs.~5(b) and 6(b). 

\section{Conclusions}      

In summary, we have formulated a method that is well suited for the analysis and 
numerical evaluation of inelastic tunneling and the determination of the nonequilibrium 
zero-frequency shot noise spectra of mesoscopic devices in the presence of a 
dispersionless phonon bath at low temperature. Following the original work of Bon\v ca 
and Trugman, we transformed the many-body EPI problem onto a one-body scattering 
problem, by projecting the original Hamiltonian in the representation of the joint 
coupled electron-phonon Fock space, i.e., the direct-product states of electron states 
and phonon number states. Based on this noninteracting Hamiltonian, we applied the 
B\"uttiker scattering theory of noise correlations in a two-terminal mesoscopic 
conductor to the case of pseudo-multi-channel leads with differing weight factors. 
Furthermore, we derived the formula for zero-frequency current-current fluctuations in 
terms of transmission and reflection amplitudes by properly taking account of the 
appropriate weight factors for the various starting channels.  

It is appropriate to remark that the present approach to EPI is based entirely on an 
approximate mapping to a one-electron picture providing that the bath is not strongly 
perturbed (if that should occur, then bath-dot many-body effects would enter the problem 
and invalidate the one-electron picture). It is believed that this approximation is 
valid only in the high bias-voltage region. Indeed, the main purpose of this paper is to 
analyze zero-frequency shot noise for inelastic tunneling under nonlinear transport 
conditions. An important advantage of this technique is that it does not involve any 
restrictions on the parameters, so that it provides a reliable solution for arbitrary 
EPI constants and tunneling strengths to the leads, as well as arbitrary internal tunnel 
couplings. 

Moreover, we have employed the derived formulae to analyze the shot noise properties of 
a single-molecular QD and for semiconductor CQDs in a series arrangement at low 
temperature. For this purpose, the NGF method provided a convenient tool to calculate 
the transmission and reflection amplitudes in the wide band approximation with the help 
of the Fisher-Lee relation. Some of the important findings of our work follow: For the 
single-molecule QD, the differential shot noise displays a peak-structure similar to 
that of the differential conductance as a function of increasing bias-voltage due to 
elastic and inelastic resonances. The Fano factor is enhanced due to EPI effects in the 
moderately high bias-voltage region. In the case of coupled QDs, both the current and 
shot noise spectra exhibit oscillatory shoulders as functions of the energy difference 
between the two QDs in the positive branch, stemming from spontaneous emission of 
phonons in nonlinear transport. In the elastic resonant tunneling region, the shot noise 
exhibits a double peak, however only a single peak shows up in the inelastic resonant 
tunneling region. The Fano factor is quite sensitive to the EPI.

\begin{acknowledgments} 

B. Dong, H. L. Cui, and N. J. M. Horing are supported by the DURINT Program administered 
by the US Army Research Office. X. L. Lei is supported by Major Projects of National 
Natural Science Foundation of China, the Special Funds for Major State Basic Research 
Project (G20000683) and the Shanghai Municipal Commission of Science and Technology 
(03DJ14003). B. Dong thanks K. Flensberg for useful email correspondence and 
discussions.

\end{acknowledgments}


\begin{thebibliography}{99}

\bibitem{Chen}{J. Chen, M. Reed, A. Rawlett, and J. Tour, Science {\bf 286}, 1550 
(1999).}

\bibitem{Liang}{W. Liang, M.P. Shores, M. Brockrath, J.R. Long, and H. Park, Nature {\bf 
417}, 725 (2002)}

\bibitem{jPark}{J. Park, A.N. Pasupathy, J.I. Goldsmith, C. Chang, Y. Yaish, J.R. Petta, 
M. Rinkoski, J.P. Sethna, H. Abruna, P.L. McEuen, and D.C. Ralph, Nature {\bf 417}, 722 
(2002).}

\bibitem{hPark}{H. Park, J. Park, A. Lim, E. Anderson, A. Allvisatos, and P. McEuen, 
Nature {\bf 407}, 57 (2000); Lam H. Yu and D. Natelson, Nano Lett. {\bf 4}, 79 (2004).}

\bibitem{Zhitenev}{N.B. Zhitenev, H. Meng, and Z. Bao, Phys. Rev. Lett. {\bf 88}, 226801 
(2002).}

\bibitem{Bose}{D. Bose and H. Schoeller, Europhys. Lett. {\bf 54}, 668 (2001); A.S. 
Alexandrov and A.M. Bratkovsky, Phys. Rev. B {\bf 67}, 235312 (2003); K.D. McCarthy, N. 
Prokof'ev, and M.T. Tuominen, Phys. Rev. B {\bf 67}, 245415 (2003).}

\bibitem{Lundin}{U. Lundin and R.H. McKenzie, Phys. Rev. B {\bf 66}, 75303 (2002).}

\bibitem{Zhu}{J.X. Zhu and A.V. Balatsky, Phys. Rev. B {\bf 67}, 165326 (2003).}

\bibitem{Flensberg}{K. Flensberg, Phys. Rev. B {\bf 68}, 205323 (2003).}

\bibitem{Braig}{S. Braig and K. Flensberg, Phys. Rev. B {\bf 68}, 205324 (2003).}

\bibitem{Mitra}{A. Mitra, I. Aleiner, and A.J. Millis, Phys. Rev. B {\bf 69}, 245302 
(2004).}

\bibitem{Fujisawa}{T. Fujisawa, T.H. Oosterkamp, W.G. van der Wiel, B.W. Broer, R. 
Aguado, S. Tarucha, Leo P. Kouwenhoven, Science {\bf 282}, 932 (1998).}

\bibitem{Brandes}{T. Brandes and B. Kramer, Phys. Rev. Lett. {\bf 83}, 3021 (1999); T. 
Brandes, N. Lambert, Phys. Rev. B {\bf 67}, 125323 (2003)}

\bibitem{Keil}{M. Keil and H. Schoeller, Phys. Rev. B {\bf 66}, 155314 (2002).}

\bibitem{Aguado}{R. Aguado and T. Brandes, Phys. Rev. Lett. {\bf 92}, 206601 (2004).}

\bibitem{Buttiker}{M. B\"uttiker, Phys. Rev. B {\bf 46}, 12485 (1992).}

\bibitem{Blanter}{For an overview of the shot noise problem, please refer to Ya.M. 
Blanter and M. B\"uttiker, Phys. Rep. {\bf 336}, 1 (2000).}

\bibitem{Dong}{B. Dong and X.L. Lei, J. Phys.: Cond. Matter {\bf 14}, 4963 (2002).}

\bibitem{Bonca}{J. Bon\v ca and S.A. Trugman, Phys. Rev. Lett. {\bf 75}, 2566 (1995).}

\bibitem{Holstein}{T. Holstein, Ann. Phys. (N.Y.) {\bf 8}, 325 (1959).}

\bibitem{Bonca2}{K. Haule and J. Bon\v ca, Phys. Rev. B {\bf 59}, 13087 (1999).}

\bibitem{Ness}{H. Ness and A.J. Fisher, Phys. Rev. Lett. {\bf 83}, 452 (1999)}

\bibitem{Emberly}{E.G. Emberly and G. Kirczenow, Phys. Rev. B {\bf 61}, 5740 (2000).}

\bibitem{Dongac}{B. Dong, H.L. Cui, and X.L. Lei, Phys. Rev. B {\bf 69}, 205315 (2004).}

\bibitem{Camalet}{S. Camalet, J. Lehmann, S. Kohler, and P. H\:a nggi, Phys. Rev. Lett. 
{\bf 90}, 210602 (2003).}

\bibitem{Fisher}{D.S. Fisher and P.A. Lee, Phys. Rev. B {\bf 23}, 6851 (1981).}

\end{thebibliography}
\end{document}